\documentclass[amsfonts,showpacs,tightenlines,aps,12pt,floatfix]{revtex4}
\usepackage{bm}
\usepackage{epsfig}
\begin{document}


\title{Models of Isospin Breaking in the Pion Form Factor:  Consequences
for the Determination of $\Pi_{\rho\omega}(m_{\rho}^2)$ and $(g-2)_{\mu}/2$}.
\author{Carl E. Wolfe}
\email[]{wolfe@yorku.ca}
\affiliation{Department of Physics and Astronomy, York University, 
4700 Keele St., Toronto, ON CANADA M3J 1P3}
\author{Kim Maltman}
\email[]{kmaltman@yorku.ca}
\affiliation{Department of Mathematics and Statistics, York University, 
4700 Keele St., Toronto, ON CANADA M3J 1P3}
\altaffiliation{CSSM, Univ. of Adelaide, Adelaide, SA 5005 AUSTRALIA}
\date{\today}

\begin{abstract}
We study the implications of several recent high-precision measurements 
of the pion form factor in the region of the $\rho-\omega$ interference
``shoulder'' for (i) the extraction of the $\rho-\omega$ mixing matrix 
element, $\Pi_{\rho\omega}(m_\rho^2)$, and (ii) the evaluation of
the isospin-breaking (IB) correction needed to incorporate hadronic
$\tau$ decay data into the determination of the 
Standard Model expectation for the leading order hadronic contribution, 
$[a_\mu]_{had}^{LO}$, to the anomalous magnetic moment of the muon,
focussing, in the latter case, on the model-dependence of the 
$\rho-\omega$ mixing component of the IB correction.
We consider a range of different models for the broad $\rho$ contribution
to the $e^+e^-\rightarrow \pi\pi$ amplitude, applying these models to
each experimental data set, and find that the model dependence 
of the $\rho-\omega$ mixing correction is significantly larger than the 
uncertainty induced by experimental errors for any individual model.
We also find that, for each such model, the recent data allows one to separate 
$\rho-\omega$ mixing and direct $\omega\to\pi\pi$ coupling
contributions to the amplitude, and hence to obtain a reasonably precise
extraction of $\Pi_{\rho\omega}(m_\rho^2)$, uncontaminated by
direct $\omega\rightarrow\pi\pi$ coupling effects, for use in
meson exchange model calculations of charge symmetry breaking
in $NN$ scattering.

\end{abstract}

\pacs{13.66.Bc,13.75.Cs,14.60.Ef,13.40.Em}

\maketitle

\section{\label{intro}Introduction}

The pion form factor, $F_\pi (s)$, 
in the $\rho$ resonance region has been the subject
of several recent experiments 
\cite{cmd203, kloe04, kloe08, snd05pipi, snd06pipi, cmd207} which, together,
provide the most precise picture yet of the structure of the $\rho$ 
resonance, including the prominent $\rho-\omega$ interference `shoulder'.  
The markedly improved precision of the new data 
(e.g. a statistical uncertainty of 1\% versus 4-5\% for earlier studies
\cite{barkov85, balakin72, benaksas72, barbiellini73}) enables high-precision
studies of isospin breaking in the $\rho-\omega$ sector and an improved 
determination of the dominant $\pi\pi$ contribution to the leading order
hadronic vacuum polarization term in Standard Model 
(SM) estimates of the anomalous
magnetic moment of the muon, $a_{\mu} \equiv (g-2)_{\mu}/2$.

Isospin breaking in the $\rho-\omega$ sector is of particular and ongoing 
importance to studies of charge symmetry breaking in the $NN$ interaction
(see Ref.~\cite{miller06} and references therein
for more details).  Significant unresolved questions include
{\it i)} the precise scale of the $\rho-\omega$ mixing matrix
element, $\Pi_{\rho\omega} \sim \langle\rho|H|\omega\rangle$,
in the region of the $\rho$ and
{\it ii)} the scale of the contribution associated with the unavoidable 
direct (i.e. non-mixing induced) isospin-breaking 
$\omega\to\pi\pi$ transition, which has often been 
neglected in determining $\Pi_{\rho\omega}$.
Previous analyses of these issues (e.g. in Ref.~\cite{mow}) were hindered 
by data of relatively lower precision, making a precise determination of
the phase between the $\rho$ and $\omega$ contributions impossible.  Without
a well-constrained phase, it is not possible to obtain a well-constrained 
estimate of the direct $\omega\to\pi\pi$ contribution, and the  
uncertainty in $\Pi_{\rho\omega}(m_{\rho}^2)$ remains large (in fact 
much greater than what is often quoted, for example in Ref.~\cite{miller06}).
The recent wealth of high-precision $e^+e^-$ data in the resonance
region represents a good first step in addressing these concerns.  
In this paper we revisit $\rho-\omega$ 
mixing in $F_\pi (s)$ and use the recent data to effect a separation 
of $\omega\to\pi\pi$ and mixing contributions, from which we determine the 
ratio of isospin-breaking (IB) to
isospin-conserving (IC) couplings, $g_{\omega_I\pi\pi}/g_{\rho_I\pi\pi}$, and 
the off-diagonal vector
meson self-energy matrix element $\Pi_{\rho\omega}(m_{\rho}^2)$.  We 
perform this analysis using a selection of models for $F_{\pi}(s)$ 
in order to clarify the numerical significance of the unavoidable model
dependence of such a separation.

One important reason for the recent interest in $F_{\pi}(s)$ is the
dominant role it plays in theoretical estimates of the 
leading order (LO) hadronic contribution to $a_{\mu}$,
$\left[ a_\mu\right]^{LO}_{had}$. $a_\mu$ is now known experimentally 
to a remarkable 0.5 parts-per-million precision as a result of BNL 
experiment E821 \cite{bnlgminus2}.
It is well known that $[a_{\mu}]_{had}^{LO}$ which,
after the purely leptonic contribution, is the
largest of the SM contributions to $a_\mu$, 
can be obtained from a weighted integral of the $e^+e^-\to hadrons$ 
cross-section~\cite{gdr69}, with the dominant contribution coming from 
the broad $\rho$ resonance.  Precise data on the $e^+e^-\to\pi^+\pi^-$ 
cross-section (or, equivalently, $F_{\pi}(s)$)
below 1 GeV is, therefore, a crucial ingredient in the determination 
of $[a_{\mu}]_{had}^{LO}$.
Use of recent $e^+e^-\to \pi^+\pi^-$ 
data as input for the $\pi\pi$ part of $[a_{\mu}]_{had}^{LO}$
leads to SM estimates for $a_{\mu}$ which  
are consistent with one another (even though the form factor measurements
themselves do not all agree on the precise $s$-dependence of 
$F_{\pi}(s)$ \cite{kloefootnote}), 
but which deviate from the experimental value by 
$\sim 
3.5-3.8\sigma$~\cite{davier2007,passera07,hmnt2007,jegerlehner09,davieretal09}.
(It should be noted, however, that 
preliminary radiative return results from BaBar differ from previous 
electroproduction results \cite{daviertau08}.)

As is well known~\cite{adh98, dh98, colangelo03, dty04, hocker04},
in the isospin limit, CVC allows the $s\leq m_\tau^2$
isovector contribution to $\left[ a_\mu\right]^{LO}_{had}$ to 
be computed using hadronic $\tau$ decay 
data~\cite{aleph97, opal99, cleo00, aleph05, dhz06, belle08taupipi} 
in place of the corresponding isovector electroproduction (EM) data.
At the present level of precision, additional small 
isospin-breaking corrections to the CVC 
relation~\cite{adh98, dh98, colangelo03, dty04, dehz02, dehz03, cen01, cen02,
gj03, fflt06, fflt07,davieretal09} 
must be taken explicitly into account. We denote the collection
of such corrections by $\left[ \delta a_{\mu}\right]_{had}^{LO}$
in what follows. Incorporating the isovector $\tau$ decay data
in this manner, one obtains estimates for the SM contribution to $a_{\mu}$ 
which deviate from the experiment by 
$\sim 1-2\sigma$~\cite{davier2007, passera07, jegerlehner09,davieretal09}.
The low-energy region from which the bulk of the SM contribution to 
$[a_{\mu}]_{had}^{LO}$ is generated is
dominated by the $\pi\pi$ component of the EM cross-section.
The IB corrections in this region, which have been extensively
studied in Refs.~\cite{cen01, cen02, gj03, fflt06, fflt07,davieretal09},
are believed to be well understood. However, the component of these 
corrections associated with $\rho-\omega$ mixing, which 
in general is model dependent, has, to date, usually been estimated 
from fits to $F_{\pi}(s)$ which employ only a single model for 
$F_{\pi}(s)$ (usually that of Refs.~\cite{cen01, cen02}). 
This procedure turns out to lead to a
significant underestimate of the resulting uncertainty,
and hence also of the uncertainty in the $\tau$ decay-based value of the 
two-pion contribution to $[a_{\mu}]^{LO}_{had}$.

In Ref.~\cite{mw06} several models were used to examine the model-dependence
of $\left[\delta a_{\mu}\right]^{LO}_{had}$. It was found that
the variation across models was significantly greater than the uncertainty
(associated with errors in the experimental data) obtained for any
individual model, though not to such an extent that it would
resolve the $\tau$ vs. $e^+e^-$ discrepancy. 
It should be borne in mind in this regard that finite energy sum
rule studies of the electromagnetic current-current correlator~\cite{kmamu},
based on the then-current EM database, yielded 
values of $\alpha_s(M_Z)\sim 0.1140-0.1150$, 
which are $\sim 1.5-2\sigma$ low compared to recent high-precision lattice 
determinations~\cite{latticealphas08}. In contrast, analogous studies
of the charged isovector current correlator~\cite{my08}, based on the
final version of the ALEPH hadronic $\tau$ decay data~\cite{aleph05},
yield a value of $\alpha_s(M_Z)$, in excellent agreement with the
new lattice results. The preliminary BaBar results for
the $e^+e^-\rightarrow \pi^+\pi^-$ cross-sections reported
by Davier at Tau'08~\cite{daviertau08}, moreover, are in better agreement
with $\tau$ expectations than with previous EM results, also
favoring the $\tau$ determination, though the $\tau$-EM
discrepancy has been reduced by the recent re-assessment of
IB corrections performed in Ref.~\cite{davieretal09}. The question of whether 
the determination of $\left[ a_\mu\right]^{LO}_{had}$ 
incorporating $\tau$ data, or that based solely on EM cross-sections, 
is the most reliable thus remains an open one. The goal of achieving
as accurate as possible an understanding of the uncertainty on 
the IB correction, $\left[ \delta a_\mu\right]^{LO}_{had}$,
to be applied to the raw $\tau$-based version of
$\left[ a_\mu\right]^{LO}_{had}$ thus remains a highly relevant one.

In the present paper, we update the analysis of the
$\rho-\omega$ mixing contribution to $\left[\delta a_\mu\right]^{LO}_{had}$
contained in Ref.~\cite{mw06}, taking into account the 
corrected version of the SND $e^+e^-\rightarrow\pi^+\pi^-$
data~\cite{snd06pipi}, the new CMD-2 $e^+e^-\rightarrow\pi^+\pi^-$
results~\cite{cmd207}, and the latest Initial State Radiation (ISR) 
determination of $F_\pi (s)$, from KLOE~\cite{kloe08}. 
In addition, we
provide details of the fit results associated with the various models
of $F_\pi (s)$, and new results for the separation of the IB part of the 
EM cross-sections into pure $\rho-\omega$ mixing and 
direct $\omega\rightarrow\pi\pi$ coupling induced contributions.
We find that, taking the model dependence of the $\rho-\omega$
contribution to $\left[\delta a_{\mu}\right]_{had}^{LO}$
into account, the uncertainty assigned to this quantity
should be increased by $1.5\times 10^{-10}$.
The improved data also allows for a more reliable 
separation of $\rho-\omega$ mixing and direct non-mixing $\omega\to\pi\pi$ 
contributions to $F_\pi (s)$.  The uncertainty in 
the scale of the non-mixing contribution, while still large due to 
weak constraints on the relative phase of the $\omega$
and $\rho$ contributions to the amplitude, is reduced by a factor of two
compared to the analysis of Ref.~\cite{mow} and favors a non-zero
value.  As a result, the central value of 
$\Pi_{\rho\omega}(m_\rho^2)$ is significantly different from that obtained
using analyses in which 
direct $\omega$ contributions are ignored.

The rest of this paper is organized as follows:  Section \ref{models} reviews 
the different models of $F_\pi (s)$ considered in our analysis;
Section \ref{data} discusses 
the four recent $e^+e^-\to\pi^+\pi^-$ data sets and the results of fits 
to each set using the models of Section \ref{models};
Section \ref{gminus2ib} updates estimates of the $\rho-\omega$ contribution
to $\left[\delta a_{\mu}\right]_{had}^{LO}$; and,
finally, Section \ref{mow} reviews the (model-dependent) separation of 
the direct $\omega\to\pi\pi$ and mixing contributions to $F_{\pi}(s)$ and 
presents the results of that analysis for the models and data sets of 
Sections \ref{models} and \ref{data}. 

\section{\label{models} Models of $F_{\pi}(s)$ in the Resonance Region}

Many recent estimates of the $\rho-\omega$ mixing contribution to 
$\left[\delta a_\mu\right]^{LO}_{had}$, which we denote by
$\left[\delta a_\mu\right]^{LO}_{had;mix}$ in what follows,
have often been based on a model of $F_{\pi}(s)$ which supplements the 
ChPT-based formulation of 
Guerrero and Pich (GP)~\cite{gp97} with a $\rho-\omega$ mixing contribution  
as proposed in Refs.~\cite{cen01, cen02} (CEN).  
We refer to this model as the GP/CEN model.  
However, $F_{\pi}(s)$ is well-described by many other models in the 
literature. Various analyses have used the K\"uhn-Santamaria (KS) 
model~\cite{ks90}, the Hidden Local Symmetry (HLS) model~\cite{hlsbasic}, 
and the Gounaris-Sakurai (GS) model~\cite{gs68}. Each model 
differs in the manner in which the 
dominant $\rho$ contribution to $F_\pi (s)$ is represented, 
with the GS and GP/CEN models explicitly taking account of 
strong final state interactions in $e^+e^-\to\pi^+\pi^-$. 
The use of models for $F_{\pi}(s)$ is unavoidable at present if one
wishes to separate the IC and IB
contributions to the EM cross-sections in the interference
region, this separation being necessary for
an estimate of $\left[\delta a_\mu\right]^{LO}_{had;mix}$.

The KS model~\cite{ks90} combines 
contributions from the $\rho$, $\rho^{\prime}$, and $\rho^{\prime\prime}$
resonances in such a way as to satisfy the charge constraint 
$F_{\pi}(s=0) = 1$. An $\omega$ contribution of the nominal
$\rho-\omega$ mixing form is added to account for the
narrow interference shoulder. 
The resulting expression for $F_\pi (s)$ is~\cite{ks90}
\begin{equation}
F_{\pi}^{\rm (KS)}(s) = \left( {\frac{P_{\rho}(s)\left (
{\frac{1 + \delta P_{\omega}(s)}{1+\delta}}\right)
+ \beta P_{\rho^{\prime}}(s) +
\gamma P_{\rho^{\prime\prime}}(s)}{1+\beta+\gamma}} \right )
\label{KSModel}
\end{equation}
where the vector meson propagator is given by 
\begin{equation}
P_V(s)  =  {\frac{m_V^2}{m_V^2-s-{\rm i}m_V\Gamma_V(s,m_V,\Gamma_V)}}\ 
\label{propagator}
\end{equation}
and $\delta$, $\beta$ and $\gamma$ are complex-valued constants.
Unless otherwise stated, $\Gamma_V(s,m_V,\Gamma_V)$ is the standard
$s$-dependent width implied by $p$-wave phase space for the 
decay of the vector meson $V$, explicitly for the $\rho$,
\begin{equation}
\Gamma_\rho(s,m_\rho,\Gamma_\rho)={\frac{\Gamma_\rho\sqrt{s}}{m_\rho}}\,
\left( {\frac{\beta (s)}{\beta (m_\rho^2)}}\right)^3
\end{equation}
where $\beta (s) = \sqrt{1-{\frac{4m_\pi^2}{s}}}$, and 
\begin{equation}
P_\omega (s)  =  {\frac{m_{\omega}^2}
     {(m_\omega^2-s-{\rm i}m_{\omega}\Gamma_{\omega})}} 
\end{equation}
with $\Gamma_{\omega}$ approximated by a constant.

An alternate version of the KS model, 
referred to in what follows as KS$^\prime$, is also sometimes employed
in the literature. The KS$^\prime$ form for $F_\pi (s)$ is
\begin{equation}
F_{\pi}^{\rm (KS^\prime)}(s) = \left( {\frac{P_{\rho}(s)\left (
1 \,+\, \delta {\frac{s}{m_\omega^2}}\, P_{\omega}(s)\right)
+ \beta P_{\rho^{\prime}}(s) +
\gamma P_{\rho^{\prime\prime}}(s)}{1+\beta+\gamma}} \right ) .
\label{KSprimeModel}
\end{equation}
It is this alternate version which is employed in Ref.~\cite{davieretal09}.
Because of strong cancellations in the region of the $\rho -\omega$
interference shoulder, the KS and KS$^\prime$ models produce significantly 
different results for $\left[\delta a_\mu\right]^{LO}_{had;mix}$.
We return to this point below. 

The HLS model~\cite{hlsbasic, hlsgood} of 
$F_\pi (s)$ has the form
\begin{equation}
F_{\pi}^{(HLS)}(s) = 1-{\frac{a_{HLS}}{2}}+{\frac{a_{HLS}}{2}}\left({\frac
{P_{\rho}(s) \left( 1 + \delta P_{\omega}(s)\right)}
{1+\delta}} \right )
\label{HLSModel}
\end{equation}
with $a_{HLS}$ a constant associated with a non-resonant 
$\gamma\pi\pi$ coupling. Despite having no explicit contribution from 
higher resonances such as the $\rho^{\prime}$, the model provides a 
good quality fit to the data below $1$ GeV. It also turns out to 
produce the correct $F_\pi (s)$ phases in the elastic $\pi\pi$
scattering region after the model parameters have been fitted~\cite{hlsgood},
though the phase constraint is not imposed in the basic structure of the model.

The GS model~\cite{gs68}, as adopted
by CMD-2 \cite{cmd203, cmd207} and used here, is given by
\begin{equation}
F_{\pi}^{\rm(GS)}(s) = {\frac{1}{(1+\beta )}}
       \left({\rm BW}_\rho^{(GS)}(s)
\left[1 + \delta{\frac{s}{m_{\omega}^2}}
P_\omega (s)\right] + 
  \beta {\rm BW}_{\rho^\prime}^{(GS)}(s)\right )
\label{GSModel}
\end{equation}
where 
\begin{equation}
{\rm BW}_V^{(GS)}(s)  =  
         {\frac {m_V^2\left(1+d(m_V){\frac{\Gamma_V}{ m_V}}\right)}
{\left(m_V^2-s+f(s,m_V,\Gamma_V)-{\rm i}m_V\Gamma_V(s,m_V,\Gamma_V)
\right)}}\label{GSprop}
\end{equation}
with
\begin{eqnarray}
d(m_V) & = & {\frac{3m_{\pi}^2}{(\pi p_{\pi}^2(m_V^2))}}\, \ell n\left(
{\frac{(m_V+2p_{\pi}(m_V^2))}{2m_{\pi}}}\right) + 
{\frac{m_V}{(2\pi p_{\pi}(m_V^2))}} -
{\frac{m_{\pi}^2 m_V}{(\pi p_{\pi}^3(m_V^2))}} \nonumber\\
f(s,m_V,\Gamma_V) & = & {\frac{\Gamma_V m_V^2}{p_{\pi}^3(m_V^2)}}\left(
   p_{\pi}^2(s)[H(s)-H(m_V^2)] + 
   (m_V^2-s)p_{\pi}^2(m_V^2){\frac{dH}{ds}}(m_V^2)\right) \nonumber\\
H(s) & = & {\frac{2 p_{\pi}(s)}{\pi\sqrt{s}}}\, \ell n\left(
     {\frac{\sqrt{s} + 2 p_{\pi}(s)}{2 m_{\pi}}}\right ) 
\end{eqnarray}
where $p_{\pi}(s) = \sqrt{{{s}\over {4}}-m_{\pi}^2}$
is the pion CM momentum for squared invariant mass $s$ and  
$\Gamma_V = \Gamma_V(m_V^2,m_V,\Gamma_V)$. 
The elastic $\pi\pi$ scattering 
phase constraint on $F_\pi (s)$ is 
reproduced by the GS model by explicit construction.

For the KS, HLS, and GS models, the constant $\delta$, which
parameterizes the strength of the narrow IB 
amplitude, is taken to be complex since a non-zero phase 
is, in general, unavoidable in the presence of an
IB direct (non-mixing) $\omega\rightarrow\pi\pi$ contribution \cite{mow}. 

Resonance chiral effective theory provides the basis for the GP model of
$F_\pi (s)$, which is given by~\cite{gp97}
\begin{equation}
F^{(GP)}_\pi (s) = 
P_\rho (s) \,  \exp\left(
{\frac{-s}{96 \pi^2 f_\pi^2}}
\left[ {\rm Re}\, 
L\left({\frac{m_\pi^2}{s}},{\frac{m_\pi^2}{m_\rho^2}}\right)\, +\, 
{\frac{1}{2}}\, {\rm Re}\, L\left( {\frac{m_K^2}{s}},
{\frac{m_K^2}{m_\rho^2}}\right)
\right]\right)\ ,
\label{gpmodel}\end{equation}
where
\begin{equation}
L\left( {\frac{m^2}{s}},{\frac{m^2}{m_\rho^2}}\right) = 
\ell n\left( {\frac{m^2}{m_\rho^2}}\right) + {\frac{8m^2}{s}} -{\frac{5}{3}}
+ \beta (s)^3 \, \ell n\left[ {\frac{\beta (s)+1}{\beta (s)-1}}\right]
\end{equation}
with $\beta (s) = \sqrt{ 1- 4m^2/s}$ and the $s$-dependent width,
$\Gamma_\rho (s,m_\rho ,\Gamma_\rho )$ appearing in $P_\rho (s)$ 
(Eq.~\ref{propagator})
replaced by the resonance chiral effective theory expression
\begin{equation}
\Gamma_\rho (s) = {\frac{m_\rho s}{96 \pi f_\pi^2}}\left(
\theta (s-4m_\pi^2)\, \beta_\pi (s)^3 \, +\,
{\frac{1}{2}} \theta (s-4m_K^2)\, \beta_K(s)^3\right)\ .
\label{gpwidth}\end{equation}
Because the GP model implements the constraints of chiral symmetry 
explicitly at next-to-leading order in the chiral expansion, 
it also ensures the correct phase for $F_\pi (s)$ 
in the elastic region to this same order.
In Ref.~\cite{cen02}, a small rescaling of the coefficient appearing on 
the RHS of Eq.~(\ref{gpwidth}) is allowed in order to account for
the $\sim 1.5$ MeV contribution of $\pi\pi\gamma$ decays
to the total width of the $\rho$~\cite{cen02,vcnote}.  We implement
such a rescaling by multiplying the RHS of Eq.~(\ref{gpwidth}) by 
\begin{equation}
1 + {\delta\Gamma_{\rho}\over\Gamma_{\rho}(m_{\rho}^2)}
\label{Grhorescale}
\end{equation}
with $\delta\Gamma_{\rho}$ a fit parameter. 
Some IB effects are incorporated
into $F^{(GP)}_\pi (s)$ if one evaluates the
phase space factors in the $s$-dependent width 
using the physical charged $\pi$ and $K$ masses.
The CEN modification of $F^{(GP)}_\pi (s)$, designed to
incorporate the $\rho-\omega$ mixing contribution not
included in the original GP model, has the form
\begin{equation}
F^{(GP/CEN)}_\pi (s) = F^{(GP)}_\pi (s) - P_\rho (s)
\left({\frac{\theta_{\rho \omega}}{3m_\rho^2}}\right)
\left({\frac{s}{m_\omega^2}}\right) P_\omega (s)\ .
\label{gpcenmodel}\end{equation}
The parameter $\theta_{\rho \omega}$ was assumed real in 
Ref.~\cite{cen02}. 

As discussed in Ref.~\cite{mw06}, the original 
version of the GP/CEN model does not provide a good fit to the 
corrected version of the earlier CMD-2 data \cite{cmd203}. 
As already noted, the fact that the $\rho-\omega$ ``mixing'' signal
is actually a combination of mixing and direct $\omega\rightarrow\pi\pi$
effects means that an effective representation of this combination 
using the form given by Eq.~(\ref{gpcenmodel}) is not generally
possible without allowing 
$\theta_{\rho \omega}$ to have a non-zero phase~\cite{mow}. 
In Ref.~\cite{mw06} the GP/CEN model was extended in this way, treating 
the phase of $\theta_{\rho\omega}$ as a fourth parameter to be fit 
to the data, and an acceptable fit became possible.  We refer to this 
version of the GP/CEN model as GP/CEN$^+$ below. In most of what follows,
however, we will focus instead on an alternate version of the GP/CEN model,
referred to below as GP/CEN$^{++}$, which
incorporates the effects of strong $\pi\pi$ rescattering not only 
in the isospin-conserving $e^+e^-\to\rho\to\pi^+\pi^-$ contribution
(the first term on the RHS of Eq.~\ref{gpcenmodel}),   
but also in the $\rho$ propagator factor of the isospin-breaking 
$\rho -\omega$ mixing contribution, leading to
\begin{equation}
F^{(GP/CEN^{++})}_\pi (s) = F^{(GP)}_\pi (s)\left ( 1 - 
\left({\frac{\theta_{\rho \omega}}{3m_\rho^2}}\right)
\left({\frac{s}{m_\omega^2}}\right) P_\omega (s)\right )\ .
\label{wmmodel}\end{equation}
where, as in the GP/CEN$^+$ version, $\theta_{\rho \omega}$ is allowed to have
a non-zero phase.  

\section{\label{data} Input Data and Fits}

Recent information on $F_\pi (s)$ in the interference region has been 
provided by the CMD-2 and SND experiments at Novosibirsk and
KLOE at DAFNE. Results are available from a total of five separate
data-taking runs. Preliminary ISR results from BaBar have also
been reported, but are not yet publicly available.

CMD-2 has reported data on $e^+e^-\rightarrow\pi^+\pi^-$ in the 
centre of mass energy range from $0.6$ to $1.0$ GeV,
taken in two separate runs, one in 1994-95~\cite{cmd203}
and one in 1998~\cite{cmd207}. We refer to these as CMD-2(94) and 
CMD-2(98), respectively. The 1998 run has five times
the luminosity of the earlier run, and correspondingly smaller statistical 
errors, but a less precise beam energy calibration.  
Whereas the beam energy was measured very precisely using resonant
depolarization for CMD-2(94), this technique was not available for 
CMD-2(98)~\cite{eidelman07} with the result that the beam energy for 
the latter is only known to a 
few parts in $10^{-3}$~\cite{cmd207}. A significant 
part of the beam energy uncertainty, moreover, 
is fully correlated \cite{eidelman07}.
The combination of increased statistics but poorer beam energy 
calibration resulted in no overall improvement in the determination of 
the $\pi\pi$ contribution to $[a_{\mu}]_{had}^{LO}$ from CMD-2(98) 
as compared to CMD-2(94)\cite{cmd207}.
Indeed, the beam energy contribution to the systematic error in 
$[a_{\mu}]_{had}^{LO}$ increased from 0.1\% in CMD-2(94) to 0.3\% 
in CMD-2(98)\cite{cmd207}.
The less precise energy calibration of CMD-2(98) also has
an impact on studies of isospin breaking in $F_{\pi}(s)$ because 
such studies require a precise experimental characterization 
of the interference shoulder.

SND results for the $e^+e^-\rightarrow\pi^+\pi^-$ cross-sections
are based on data taken from 1996-2000 with centre of mass energies in the
range $0.39$ to $0.97$ GeV~\cite{snd05pipi, snd06pipi}. 
The SND data carries a slightly larger systematic error than CMD-2(94), but
much smaller statistical uncertainty and slightly better beam energy 
calibration.  The CMD-2 and SND data sets are in good overall 
agreement.

The most recent, KLOE, measurements of the
$e^+e^-\rightarrow\pi^+\pi^-$ cross-section
are obtained using the ISR technique~\cite{kloe04, kloe08}. 
It is well known that the earlier, 2004, KLOE results
for $F_\pi (s)$~\cite{kloe04}, based on data collected in 2001,
deviated in shape from both the CMD-2 and SND results,
lying systematically higher than either below $\sqrt{s} = 700$ MeV, and 
systematically lower beyond this~\cite{snd05pipi, cmd207}.  We refer to 
these results as KLOE(01) below.
A new analysis has recently been completed using data 
collected in 2002 and an improved analysis strategy~\cite{kloe08}. 
The new data represents twice as many events as before and 
corresponds to an integrated luminosity that is 1.7 times 
greater, and the analysis yields significantly reduced 
statistical and systematic uncertainties, with the systematic
uncertainties now dominant.
In addition, part of the previous shape discrepancy has been
resolved, KLOE now being in better agreement with CMD-2(98) below 
the $\rho$ peak. A reduced disagreement persists above the $\rho$ peak.
In view of the fact that the results of the new analysis
supercede those of KLOE(01)~\cite{kloe08},
our study of IB in the interference region, reported below, 
is based on the former, which we refer to for clarity
as KLOE(02) in what follows. 

Our earlier analysis of $\left[\delta a_{\mu}\right]^{LO}_{had;mix}$, 
reported in Ref.~\cite{mw06}, was based on the CMD-2(94) and uncorrected
SND data sets. The KLOE(01) results were also used, though, without full 
covariance information, the resulting fits were unreliable. The fits 
presented below 
reflect the new CMD-2(98) data, the corrected SND data, and the new
KLOE(02) data, including full covariance information, where available.

We have performed fits to each data set using the models described
previously in Section~\ref{models}.  In all cases the results shown below correspond to the bare 
form factor, that is the form factor with the effects of vacuum 
polarization
removed \cite{vpfootnote, footnote2}.  Where applicable, the following input 
values have been used:  
$m_{\omega} = 782.68$ MeV, 
$\Gamma_{\omega} = 8.68$ MeV,
$m_{\rho^{\prime}} = 1465$ MeV \cite{footnote3},
$\Gamma_{\rho^{\prime}} = 400$ MeV (310 MeV for the GS model \cite{GSrhopfootnote}),
$f_{\pi} = 92.4$ MeV,
$m_{\pi} = 139.57$ MeV, and
$m_K = 495$ MeV.
Although a contribution from the $\rho^{\prime\prime}$ appears in the KS model,
we have found that, because the data we consider 
is limited to $\sqrt{s}$ below 1 GeV, it
provides insufficient constraints on this term.  We manually set the 
$\rho^{\prime\prime}$ contribution to zero in the KS model to avoid it being 
used by the minimization algorithm to `fine tune' the $\rho$ region of the 
fit without regard for the consequences at higher $\sqrt{s}$. 
For similar reasons the coefficient of the $\rho^{\prime}$ contribution, 
$\beta$, is taken to be real.
Fit parameters and their covariances were obtained using MINUIT.
The results for each model and for each data set are shown in 
Tables \ref{table1}, \ref{table2}, \ref{table3}, and \ref{table4}.  
A blank entry indicates that a fit parameter is inapplicable 
to that particular model.  For the GP/CEN$^+$ and GP/CEN$^{++}$ 
models, the effective
value of $\Gamma_{\rho}$ is shown in brackets to highlight that it is 
in fact $\delta\Gamma_{\rho}$ which is the fit parameter.  Note also that for 
both of these models the quantity $|\theta_{\rho\omega}|$ is replaced by 
$|\delta| \equiv |\theta_{\rho\omega}|/3m_{\rho}^2$ in the tables
to facilitate comparison with the other models.

\begin{table}
\caption{\label{table1}
Results of fits to the CMD-2(94) data.}
\vskip .1in
\begin{tabular}{|c||c|c|c|c|c|}
\hline
Parameter & KS & HLS & GS & GP/CEN$^+$ & GP/CEN$^{++}$ \\
\hline
$m_{\rho}$ (MeV) & 772.42$\pm$0.63 & 773.84$\pm$0.62 & 774.67$\pm$0.64 & 775.89$\pm$0.61 & 775.89$\pm$0.61 \\
$\Gamma_{\rho}$ (MeV) & 140.47$\pm$1.34 & 143.89$\pm$1.52 & 144.37$\pm$ 1.43 & (145.51) & (145.47) \\
$\delta\Gamma_{\rho}$ (MeV) & - & - & - & -1.83$\pm$0.63 & -1.86$\pm$ 0.63\\
$|\delta|$ ($10^{-3}$) & 1.77$\pm$0.14 & 1.84$\pm$0.14 & 1.84$\pm$0.14 & 2.25$\pm$0.17 & 1.91$\pm$ 0.14 \\
${\rm Arg}(\delta)$ (deg) & 14.9 $\pm$ 3.3 & 14.1 $\pm$ 3.2 & 14.3 $\pm$ 3.2 & 14.1 $\pm$ 2.9 & 14.2 $\pm$ 2.9 \\
$\beta$  & -0.136$\pm$0.005 & - & -0.073$\pm$0.005& - & - \\
$a_{HLS}$ & - & 2.3479$\pm$0.016 & - & - & - \\
\hline
$\chi^2$/dof & 37.1/38 & 36.6/38 & 35.8/38 & 40.5/39 & 40.6/39 \\
\hline
\end{tabular}
\end{table}

\begin{table}
\caption{\label{table2}
Results of fits to the CMD-2(98) data.}
\vskip .1in
\begin{tabular}{|c||c|c|c|c|c|}
\hline
Parameter & KS & HLS & GS & GP/CEN$^+$ & GP/CEN$^{++}$ \\
\hline
$m_{\rho}$ (MeV) & 772.65$\pm$0.55 & 774.07$\pm$0.54 & 774.87$\pm$0.55 & 776.86$\pm$0.51 & 776.86 $\pm$ 0.52 \\
$\Gamma_{\rho}$ (MeV) & 141.56$\pm$0.81 & 146.51$\pm$1.01 & 146.61$\pm$ 0.91 & (147.66) & (147.63) \\
$\delta\Gamma_{\rho}$ (MeV) & - & - & - & -0.32$\pm$0.56 & -0.34$\pm$ 0.56\\
$|\delta|$ ($10^{-3}$) & 1.88$\pm$0.068 & 1.94$\pm$0.069 & 1.93$\pm$0.069 & 2.36$\pm$0.08 & 2.00$\pm$ 0.069 \\
${\rm Arg}(\delta)$ (deg) & 15.1 $\pm$ 3.5 & 11.2 $\pm$ 3.9 & 12.2 $\pm$ 3.5 & 12.5 $\pm$ 3.1 & 12.6 $\pm$ 3.1 \\
$\beta$  & -0.137$\pm$0.003 & - & -0.079$\pm$0.003& - & - \\
$a_{HLS}$ & - & 2.3690$\pm$0.011 & - & - & - \\
\hline
$\chi^2$/dof & 34.4/24 & 18.1/24 & 18.6/24 & 39.1/25 & 39.4/25 \\
\hline
\end{tabular}
\end{table}

\begin{table}
\caption{\label{table3}
Results of fits to the SND data.}
\vskip .1in
\begin{tabular}{|c||c|c|c|c|c|}
\hline
Parameter & KS & HLS & GS & GP/CEN$^+$ & GP/CEN$^{++}$ \\
\hline
$m_{\rho}$ (MeV) & 771.60$\pm$0.46 & 773.42$\pm$0.45 & 774.26$\pm$0.47 & 776.37$\pm$0.44 & 776.37$\pm$0.44 \\
$\Gamma_{\rho}$ (MeV) & 142.77$\pm$0.78 & 147.23$\pm$0.89 & 147.34$\pm$ 0.84 & (147.34) & (147.31) \\
$\delta\Gamma_{\rho}$ (MeV) & - & - & - & -0.32$\pm$0.42 & -0.35$\pm$ 0.42\\
$|\delta|$ ($10^{-3}$) & 1.96$\pm$0.07 & 2.04$\pm$0.07 & 2.02$\pm$0.07 & 2.47$\pm$0.08 & 2.09$\pm$ 0.07 \\
${\rm Arg}(\delta)$ (deg) & 16.2 $\pm$ 1.8 & 15.1 $\pm$ 1.7 & 15.4 $\pm$ 1.7 & 15.8 $\pm$ 1.6 & 15.9 $\pm$ 1.6 \\
$\beta$  & -0.144$\pm$0.003 & - & -0.081$\pm$0.003& - & - \\
$a_{HLS}$ & - & 2.3770$\pm$0.0098 & - & - & - \\
\hline
$\chi^2$/dof & 34.5/40 & 23.6/40 & 25.1/40 & 62.6/41 & 62.9/41 \\
\hline
\end{tabular}
\end{table}

\begin{table}
\caption{\label{table4}
Results of fits to the KLOE(02) data.}
\vskip .1in
\begin{tabular}{|c||c|c|c|c|c|}
\hline
Parameter & KS & HLS & GS & GP/CEN$^+$ & GP/CEN$^{++}$ \\
\hline
$m_{\rho}$ (MeV) & 771.21$\pm$0.18 & 773.05$\pm$0.17 & 773.90$\pm$0.18 & 776.91$\pm$0.16 & 776.91$\pm$0.16 \\
$\Gamma_{\rho}$ (MeV) & 141.88$\pm$0.36 & 147.29$\pm$0.41 & 147.04$\pm$ 0.39 & (146.50) & (146.46) \\
$\delta\Gamma_{\rho}$ (MeV) & - & - & - & -2.29$\pm$0.21 & -2.33$\pm$ 0.21\\
$|\delta|$ ($10^{-3}$) & 1.26$\pm$0.08 & 1.49$\pm$0.08 & 1.48$\pm$0.08 & 2.32$\pm$0.10 & 1.98$\pm$ 0.08 \\
${\rm Arg}(\delta)$ (deg) & 19.4 $\pm$ 7.4 & 6.0 $\pm$ 6.3 & 9.4 $\pm$ 6.3 & 23.2 $\pm$ 4.5 & 23.5 $\pm$ 4.5 \\
$\beta$  & -0.145$\pm$0.0011 & - & -0.085$\pm$0.0011& - & - \\
$a_{HLS}$ & - & 2.3880$\pm$0.0034 & - & - & - \\
\hline
$\chi^2$/dof & 293.2/55 & 63.6/55 & 72.4/55 & 315/56 & 318/56 \\
\hline
\end{tabular}
\end{table}

It is apparent from Tables~\ref{table1}, \ref{table2}, \ref{table3}, and
\ref{table4} that, for each given model, all four data sets are in broad 
agreement with each other.  The different models are also in general 
agreement with each other for each given data set.  One exception is 
the GP/CEN$^+$ model, which yields a value of $|\delta|$ 
$\sim 20\%$ larger than in the other models. This deviation,
however, disappears when the effects of rescattering are incorporated
into the $\rho$ propagator factor which enters the $\rho-\omega$
interference contribution through use of the modified 
GP/CEN$^{++}$ form.  
Table~\ref{table4} reveals that fits to the KLOE data stray most 
from the fits to the other three data sets, in particular through 
substantially different central values for the magnitude and phase of the 
isospin-breaking parameter $\delta$, and a much greater 
model-dependence in the value of the phase.
Note that, in contrast to the situation for
the other data sets, only the HLS and GS models provide 
acceptable quality fits to the KLOE data. 
That the 
phase error is so much greater for the KLOE(02) fits than for any of the other 
data sets is likely a reflection of the small number of data points 
(five) in the interference region ($770-800$ MeV) compared to 
CMD-2(94), CMD-2(98), and SND which reported, respectively, 10, 12, and 
13 data points in that range and so provide a significantly 
better characterization of the interference shoulder.

Looking across models within each data set, we find that the 
HLS, GS, GP/CEN$^+$, and GP/CEN$^{++}$ models
favour a slightly larger $\rho$ mass 
than does the KS model.  
For comparison, Belle has recently reported the analogous
values for the charged $\rho$ parameters,
$m_{\rho^\pm} = 774.6\pm 0.5 \,{\rm MeV}$ and
$\Gamma_{\rho^\pm} = 148.1\pm 1.7\,{\rm MeV}$, obtained 
from $\tau^\pm\rightarrow \pi^\pm \pi^0\nu_\tau$
using the GS model for the $\rho$, $\rho^\prime$ and 
$\rho^{\prime\prime}$ Breit-Wigner shapes~\cite{belle08taupipi}. 

With three exceptions, the quality of the fits, as measured by the 
$\chi^2$/dof, is reasonable.  
The fits to the KLOE(02) data using the KS, GP/CEN$^+$ and
GP/CEN$^{++}$ models yield unacceptably large
$\chi^2$/dof, and hence are not included when 
arriving at a final combined assessment of our results below.
The GS and HLS models provide the best quality fits in
all cases, and the only acceptable quality fits to the KLOE data.
The fit residuals (not shown here) reveal that the GP/CEN$^+$ and GP/CEN$^{++}$ 
models tend to lie below the data above $\sqrt{s} = m_\rho$, 
which suggests that,
with the high quality of the current data sets, forms of the resonance chiral 
effective theory model which include explicit higher resonance 
(in particular $\rho^\prime$) contributions will be needed if one wants to 
further improve the fit quality.

\section{\label{gminus2ib} Implications for $a_{\mu}$}

As is well known, the leading order
hadronic vacuum polarization contribution to $a_{\mu}$ can be obtained 
from the experimental $e^+e^-\to hadrons$ cross section using the 
dispersive representation~\cite{gdr69}
\begin{equation}
[a_{\mu}]_{had}^{LO} = {\alpha_{EM}^2(0)\over 3\pi^2}\int_{4m_{\pi}^2}^\infty ds{K(s)\over s}R(s)
\label{kernel}
\end{equation}
where $K(s)$ has the form given in \cite{gdr69} and $R(s)$ is the ratio 
of the ``bare'' $e^+e^-\to hadrons$ cross-section to that for 
$e^+e^-\to \mu^+\mu^-$.
The integrand in Eq.~\ref{kernel}, in terms of $F_\pi (s)$,
has the form $R(s) = (1-4m_\pi^2/s)^{3/2}|F_\pi(s)|^2/4$. 
Fitting the data using any of the models for $F_{\pi}(s)$ 
discussed above, in which the IB contribution is that associated with the 
$\omega$, thus allows a straightforward, albeit model-dependent,
separation of the $\rho$-$\omega$ IB and nominally IC components 
of the experimental cross-sections. 
Writing $F_\pi (s)=F_{\pi ,NIC}(s)+F_{\pi ,mix}(s)$, with
$F_{\pi , mix}(s)$ and $F_{\pi ,NIC}(s)$ the $\rho$-$\omega$ IB and
nominally IC (NIC) ($\rho +\rho^\prime +\cdots$) contributions to $F_\pi (s)$,
and defining $\delta|F_\pi(s)|^2\equiv |F_\pi(s)|^2-|F_{\pi ,NIC}(s)|^2$,
we obtain the desired $\rho-\omega$ mixing 
plus direct $\omega\rightarrow\pi\pi$ coupling contribution 
to $\left[\delta a_\mu\right]^{LO}_{had}$,
\begin{equation}
[\delta a_{\mu}]_{had;mix}^{LO} = {\alpha_{EM}^2(0)\over 12\pi^2}
\int_{4m_{\pi}^2}^{s_{max}}ds{K(s)\over s}(1-4m_\pi^2/s)^{3/2}\delta|F_\pi(s)|^2.
\label{IBvpol}
\end{equation}
This represents one component of the full IB correction which must be
applied to the experimental $\tau\rightarrow\pi\pi\nu_\tau$ data, 
the other components being 
those associated with (i) the impact of the $\pi^\pm-\pi^0$ mass difference on 
the phase space factor, (ii) possible differences in the $\rho^0$ and 
$\rho^\pm$ masses and widths, 
(iii) possible differences in the products of decay constants and
$\pi\pi$ couplings of the various charged and neutral $\rho$ resonances, and 
(iv) long-distance electromagnetic 
corrections~\cite{cen01,cen02,gj03,fflt06,fflt07}.
Many, though not all, of recent $\tau$-based analyses of 
$\left[ a_\mu\right]^{LO}_{had}$~\cite{adh98, dh98, colangelo03, dty04, 
dehz02, dehz03, hocker04, jegerlehner09} 
employ a value of $[\delta a_{\mu}]_{had; mix}^{LO}$ 
obtained using the GP/CEN model \cite{cen01, cen02}.  
In Ref.~\cite{mw06} it was pointed out that the generic structure
of the $\rho-\omega$ interference contribution to $F_{\pi}(s)$ and the 
monotonically decreasing nature of the integration kernel, $K(s)$, combine 
to introduce strong fit-parameter-sensitive cancellations in the integral 
in Eq.~\ref{IBvpol}, and hence 
significant model dependence into $[\delta a_{\mu}]_{had;mix}^{LO}$
(see below for more on this point). The variation in the values 
of $[\delta a_{\mu}]_{had;mix}^{LO}$ across the various
models was found to be greater than the experimental 
uncertainty produced by any single model, and to necessitate
a re-evaluation of the overall uncertainty on this quantity.

\begin{table}
\caption{\label{amutable}
$[\delta a_{\mu}]^{LO}_{had;mix}\times 10^{10}$ for the models discussed in the text and
the CMD-2, SND, and KLOE $e^+e^-\to\pi^+\pi^-$ cross-sections.}
\vskip .1in
\begin{tabular}{|c||c|c|c|c|c|}
\hline
Experiment & KS & HLS & GS & GP/CEN$^+$ & GP/CEN$^{++}$ \\
\hline
CMD-2(94) & $3.8\pm 0.6$ & $4.0\pm 0.6$ & $2.0\pm 0.5$ & $2.0\pm 0.5$ & $1.8\pm 0.4$ \\
CMD-2(98) & $4.0\pm 0.6$ & $4.6\pm 0.6$ & $2.5\pm 0.5$ & $2.2\pm 0.4$ & $2.1\pm 0.4$ \\
SND & $4.2\pm 0.4$ & $4.3\pm 0.4$ & $2.2\pm 0.3$ & $1.9\pm 0.3$ & $1.7\pm 0.3$ \\
KLOE(02) & ($2.2\pm 0.6$) & $4.2\pm 0.7$ & $2.2\pm 0.6$ & ($0.5\pm 0.8$) & ($0.3\pm 0.8$) \\
\hline
\end{tabular}
\end{table}

Updating the analysis of Ref.~\cite{mw06} to incorporate the data 
improvements described in Section~\ref{data}, 
we obtain the values of $[\delta a_{\mu}]_{had; mix}^{LO}$ shown in 
Table~\ref{amutable} for the models described above 
(with $s_{max} = 2.25\; {\rm GeV}^2$).
The results for the KLOE(02) data + (KS, GP/CEN$^+$,
GP/CEN$^{++}$) model cases are shown in brackets to highlight the fact 
that the underlying fits are of poor quality 
and that these values are not used in fixing any of the central values,
or error ranges, quoted below. 

We see that, despite the higher statistics of the CMD-2(98) data,
the uncertainty in $[\delta a_{\mu}]^{LO}_{had;mix}$ has 
not been reduced. This is because
of the increased uncertainty in the phase of the IB term 
resulting from the poorer beam energy calibration.
The corrections to the original SND data (the
SND results quoted in Ref.~\cite{mw06} were based on
the uncorrected results) turn out to have little 
impact on $[\delta a_{\mu}]^{LO}_{had;mix}$.
The introduction of the new KLOE(02) data and 
covariance information allows us to significantly improve
on our earlier treatment of the now-superceded KLOE(01) results.
We note, in particular, that the KLOE(02)-based values for the HLS and
GS models are now consistent with those obtained from the other two 
experiments.  

In arriving at a final assessment of our results for
$\left[\delta a_\mu\right]^{LO}_{had;mix}$, we have adopted the
view that, since all the models considered have a
reasonable basis in phenomenology, all results corresponding
to a given data set and given model which produce an
acceptable quality fit are to be included in the assessment.
Our final results (both here, and for the quantities to be
discussed in the next section) are thus obtained
by first performing a weighted average over all experiments
for each separate model, and then taking the average (half the
difference) of the maximum and minimum values allowed by the resulting
error intervals for the different models to define our central values 
(model-dependence-induced uncertainties).
The result of this prescription for 
$\left[\delta a_{\mu}\right]^{LO}_{had;mix}$ is
\begin{equation}
\label{CVall}
\left[\delta a_{\mu}\right]^{LO}_{had;mix} = (3.1\pm 1.5_{model}\pm 0.3_{data})\times 10^{-10},
\end{equation}
whose central value is compatible with that reported in Eq.~15 of 
Ref.~\cite{mw06}.  

Two features of the analysis worth emphasizing
are the strong sensitivity of $\left[\delta a_{\mu}\right]^{LO}_{had;mix}$
to the phase, $\phi$, of the IB contribution 
and the resulting sensitivity to the model
employed for the broad $\rho$ contribution. Both can be understood
by considering the generic form
of the flavor `38' component of the EM cross-section. 

Writing the isospin-breaking amplitude in the ``$\rho-\omega$ mixing" 
form, generically $\propto B_\rho(s)\delta P_\omega(s)$, with 
$\delta = |\delta|e^{i\phi}$ and $B_\rho(s)$ the 
model-dependent form of the $\rho (770)$ Breit-Wigner,
the IB, flavor `38' component of the EM 
cross-section is $\propto$
\begin{equation}
\label{genericmixing}
|B_\rho(s)|^2\left[{2|\delta|m_\omega^2 \over \left[(m_\omega^2-s)^2
+ m_\omega^2\Gamma_\omega^2\right]}\right] \bigl(\cos(\phi)(m_\omega^2-s) -
m_\omega\Gamma_\omega\sin(\phi)\bigr).
\end{equation}
The coefficient multiplying $\cos(\phi)$ is antisymmetric in $s$ about 
$m_{\omega}^2$, which means that the contribution of this term to 
$[\delta a_{\mu}]^{LO}_{had;mix}$ vanishes in the limit that the 
$s$-dependence of $B_\rho(s)K(s)/s$ is neglected. The result is a 
relative enhancement of the contribution of the $\sin(\phi)$ term,
and so a significant reduction of $[\delta a_{\mu}]^{LO}_{had;mix}$ 
relative to what would be obtained for $\phi=0$,
for even modest values of $\phi$. Thus, relaxing the $\phi =0$
assumption of the GP/CEN model, the non-zero phase $\phi$
preferred by the data for the GP/CEN$^+$ and GP/CEN$^{++}$ models 
leads to the reduction seen in Table~\ref{amutable}
as compared to the GP/CEN result 
$[\delta a_{\mu}]^{LO}_{had;mix}=(3.5\pm0.6)\times 10^{-10}$,
which was obtained from the uncorrected CMD-2(94) data, and $\phi$
fixed to $0$~\cite{cen02, dehz02, footnote10}.  

Regarding the sensitivity to the model chosen for the form
of the $\rho$ contribution, the $\sim 1.5\times 10^{-10}$
model-dependence uncertainty noted above is
significantly greater than the $\sim 0.5\times 10^{-10}$
uncertainty generated by experimental errors for any given
model. For a given model, however, the different data sets produce
results which are in good agreement.

Before continuing, it is worth commenting on one feature of
the relation of our results to those of Ref.~\cite{davieretal09}.
The latter reference, employing the KS$^\prime$ model, finds results for
$\left[\delta a_\mu\right]_{had;mix}^{LO}$ compatible with those
obtained using the GS model, in contrast to our results above.
The reason for this difference is that we employ the KS model,
whose $s$-dependence in the region of the interference shoulder
is slightly different from that of the KS$^\prime$ model. We have
verified that the KS$^\prime$ model indeed produces results
compatible with those of the GS model, as found in
Ref.~\cite{davieretal09}, the values corresponding to
the CMD2(94), CMD2(98) and SND(06) data sets
being $2.1(5)\times 10^{-10}$, $2.2(6)\times 10^{-10}$ and 
$2.3(3)\times 10^{-10}$, respectively. The difference between
the KS and KS$^\prime$ model results serves as a further illustration 
of our point about the model sensitivity of the determination of
$\left[\delta a_\mu\right]_{had;mix}^{LO}$. This sensitivity
is a consequence of the strong cancellation in the integral
for $\left[\delta a_\mu\right]_{had;mix}^{LO}$, a cancellation
which enhances the impact of small differences in the
$s$-dependence of the $F_{\pi, mix}(s)$ among the different models.

With the KS model form, the IB parameter $\delta$ also occurs
in the nominally IC $\rho$ contribution to the amplitude. One might think
it appropriate to also remove the effect of this IB contribution
in comparing the $\tau$ and electroproduction results. Such a
procedure, however, would involve double counting 
when used in combination with a separate assessment (such as that 
discussed in Refs.~\cite{cen01,cen02}) of the IB contribution 
to the broad $\rho$ component
of the ratio of charged and neutral current versions of $F_\pi$. 
The reason is that any $\delta$-dependence in 
$F_{\pi ,NIC}$ represents only a partial assessment of IB in the coefficients
$1/[(1+\delta )(1+\beta +\gamma )]$, $\beta /(1+\beta +\gamma )$ and 
$\gamma /(1+\beta +\gamma )$ of the $\rho ,\, \rho^\prime ,\, 
\rho^{\prime\prime}$ contributions to $F_\pi (s)$, coefficients
which are proportional to the products
$f_\rho g_{\rho \pi\pi},\, f_{\rho^\prime} g_{\rho^{\prime} \pi\pi},\, 
\cdots$ of the various $\rho$ resonance decay constants and $\pi\pi$ 
couplings. An assessment of IB contributions to these products 
is, however, already 
provided by the treatment of Refs.~\cite{cen01,cen02}. 

While the $\rho -\omega$ interference shoulder, 
because of its narrow structure, is unambiguously 
identifiable as an IB effect associated with the intermediate $\omega$ state,
there is no signal, internal to the electroproduction
data, that allows one to identify IB contributions to the
broad $\rho ,\, \rho^\prime ,\, \rho^{\prime\prime}$ parts
of the cross-section. Thus while, for the KS model, the presence of
the IB parameter $\delta$ in the broad $\rho$
contribution produces an explicit IB contribution to 
$f_\rho g_{\rho \pi\pi}$, additional implicit IB contributions are, 
in general, also unavoidably present in the coefficients 
$1/(1+\beta +\gamma )$, $\beta /(1+\beta +\gamma )$ and 
$\gamma /(1+\beta +\gamma )$. Indeed, if one 
simply reparametrizes the KS model, writing
\begin{equation}
F_{\pi}^{\rm (KS)}(s) = \left( {\frac{P_{\rho}(s)\left (
1 + \delta P_{\omega}(s)\right)
+ \beta^\prime P_{\rho^{\prime}}(s) +
\gamma^\prime P_{\rho^{\prime\prime}}(s)}{1+\delta +
\beta^\prime +\gamma^\prime}} \right ) ,
\label{KSModelalt2}
\end{equation}
explicit IB dependence, which
in the original parametrization was present only in the $\rho$
and $\rho -\omega$ ``mixing'' contributions, is now present
in all of the $\rho$, $\rho^\prime$, $\rho^{\prime\prime}$
and $\rho -\omega$ ``mixing'' contributions. The presence or
absence of such explicit IB in the nominally
IC $\rho ,\, \rho^\prime ,\, \rho^{\prime\prime}$ contributions
to $F_\pi (s)$ is thus parametrization-dependent, from which it
follows that the integrated contribution associated with such 
explicit IB contributions for any particular parametrization
has no well-defined physical meaning. 

While framed in the context
of the KS model, the above discussion, of course, applies more generally;
implicit IB contributions, which cannot be identified experimentally, 
will also be present in the NIC part of the fitted version of 
$F_\pi$ obtained using any model. Fortunately, there is no need
to tackle the difficult task of trying to identify
such additional implicit IB contributions;
Refs.~\cite{cen01,cen02} provide a theoretically
well-founded framework for assessing the full set of 
IB contributions to the NIC part of $F_\pi$, one that is expected 
to produce reliable results in the region
below $s\sim 1\ {\rm GeV^2}$ which dominates the SM contributions to
$\left[ a_\mu\right]^{LO}_{had}$. With sufficiently
good $\tau$ decay and electroproduction data one could, of
course, instead, fit any particular model separately to both data sets and 
identify small IB effects in the fitted NIC contributions
that result. Current data is insufficiently precise to make such an 
experimental alternative feasible at present. 

We note finally that, independent of the above discussion,
one could create a third variant of ``the'' KS model by replacing
the factor $s/m_\omega^2$ in the mixing term of the KS$^\prime$
model with a function $f(s)$ which varies smoothly from $0$
at $s=0$ to $1$ in the region around the $\omega$. Such a variant
would share with the KS$^\prime$ version of the model the absence of explicit
IB in $F_{\pi ,NIC}(s)$, but produce a numerical result for the
$\rho -\omega$ ``mixing'' correction matching that of the KS version.

As previously mentioned, the $\rho-\omega$ mixing contribution is a 
part of the ratio of EM to weak form factors, which is itself one of three
components of the overall IB correction, $R_{IB}$, to the
$\tau$-based determination of $\left[\delta a_\mu\right]^{LO}_{had}$
considered in Ref.~\cite{cen02}.
The uncertainty in the contribution of the ratio of form factors also 
happens to be the dominant
uncertainty in the total correction $\left[\delta a_\mu\right]_{had}^{LO}$
associated with $R_{IB}$.  The results presented in Eq.~\ref{CVall}
show that the 
$\rho-\omega$ contribution to the form factor ratio 
has a significant model-dependence, necessitating an increase 
of $\sim 1.5\times 10^{-10}$ in the overall uncertainty assigned to
$\left[\delta a_\mu\right]_{had}^{LO}$.
We argue that such model dependence should be added linearly
to the remaining uncertainty to reflect the range that would be obtained
by using different phenomenologically acceptable model fits.
The overall uncertainty in the ratio of form factors 
thus shifts up to $\pm 3.9\times 10^{-10}$ if one uses 
the figures of Ref.~\cite{dehz02} for the other IB corrections, 
and up to $\pm 3.1\times 10^{-10}$ if one uses instead the
more recent results of Ref.~\cite{davieretal09} for these corrections.
Modifying the $\tau$-based assessment of Ref.~\cite{davieretal09}
to account for both the slight shift represented by our version of the central
value and our increased model-dependence of
$\left[\delta a_\mu\right]^{LO}_{had;mix}$, we obtain the following
modified version of the difference between the experimental and
$\tau$-based SM assessment of $a_\mu$,
\begin{eqnarray}
a_\mu^{exp}-a_\mu^{\tau}=(14.5\pm 9.7)\times 10^{-10}
\end{eqnarray}
where, as above, we have added the contribution associated with the
increased model-dependence of $\left[\delta a_\mu\right]^{LO}_{had;mix}$ 
linearly to the other errors.

\section{\label{mow} Extraction of $\Pi_{\rho\omega}$ and the Direct 
$\omega\to\pi\pi$ Coupling}

\subsection{Analysis}

In general it is unavoidable that, for a given choice of $\rho$
and $\omega$ interpolating fields,  
the $\rho-\omega$ interference signal in $F_\pi (s)$ will be
the result of a combination of true $\rho-\omega$ mixing 
and a direct (non-mixing-induced) IB $\omega\to\pi\pi$ contribution.
It is of interest to 
separate these contributions in order to gauge the 
strength of the intrinsic $\omega\pi\pi$ coupling and to estimate 
the off-diagonal $\rho-\omega$ element, $\Pi_{\rho\omega}(q^2)$,
of the vector meson self-energy matrix. Such a separation is 
relevant for meson-exchange models of IB in the $NN$ interaction, 
which models require information about $\Pi_{\rho\omega}(q^2)$ 
at $q^2<0$. Often an ``effective'' $\Pi_{\rho\omega}(q^2)$,
extracted from electroproduction data under the assumption that
direct $\omega\rightarrow\pi\pi$ contributions can be neglected,
has been used for this purpose (see, e.g., Ref.~\cite{cb87}).
In general, such
an estimate will be
contaminated by direct $\omega\rightarrow\pi\pi$ contributions,
which are unavoidably present in the electroproduction 
cross-section in the interference region.

The separation of mixing and direct $\omega\rightarrow\pi\pi$ 
contributions depends on the model used to represent the broad 
$\rho$ contribution to $F_{\pi}(s)$. The feasibility of such a 
separation was discussed in detail 
in Ref.~\cite{mow}, though the lower quality of the data then available 
\cite{barkov85} allowed only weak constraints to be placed on 
$\Pi_{\rho\omega}$ and the strength of the direct IB
$\omega\to\pi\pi$ coupling, $g_{\omega^I\pi\pi}$.
Below, we revisit this analysis using the recent 
data described in section \ref{data} and the various 
models of $F_{\pi}(s)$ described in section~\ref{models}. 

In the interference region, the time-like EM pion form factor is given,
in terms of the {\it physical} vector meson degrees of freedom, by 
\begin{equation}
F_{\pi}(s) = \left[g_{\rho\pi\pi}D_{\rho\rho}(s){f_{\rho\gamma}\over e} +
g_{\omega\pi\pi}D_{\omega\omega}(s){f_{\omega\gamma}\over e}\right] +
{\rm background}
\label{ff1}\end{equation}
where $D_{VV}(s)$ are the scalar propagators of the physical vector 
mesons, $g_{\rho\pi\pi}$ and $g_{\omega\pi\pi}$ are the couplings of the 
physical
$\rho$ and $\omega$ to the two-pion final state, and $f_{\rho\gamma}$ and 
$f_{\omega\gamma}$ are the electromagnetic $\rho$ and $\omega$ couplings.
The background term includes all non-resonant contributions to the form 
factor.

The physical $\rho$ and $\omega$ fields referred to in Eq.~(\ref{ff1}) are
admixtures of the isospin-pure $I=1$ and $I=0$ fields and can be written as
\begin{equation}
\rho = \rho^I-\epsilon_1\omega^I, \;\;\;\; \omega = \omega^I+\epsilon_2\rho^I ,
\label{mix1}\end{equation}
with $\epsilon_1$ and $\epsilon_2$ as given in Ref.~\cite{mow} to first order 
in isospin breaking.
The physical $\omega\rightarrow\pi\pi$ coupling, $g_{\omega\pi\pi}$, 
thus contains both a mixing contribution and a direct 
IB coupling contribution.  Explicitly~\cite{mow}, 
\begin{equation}
g_{\omega\pi\pi} = g_{\omega^I\pi\pi}(1-z)-iz\tilde{T}g_{\rho^I\pi\pi}
\label{coupling}
\end{equation}
where 
\begin{eqnarray}
z &\equiv& \left [ 1 -
{\hat{m}_{\omega}\Gamma_{\omega}\over \hat{m}_{\rho}\Gamma_{\rho}} -
i\left ({\hat{m}_{\omega}^2-\hat{m}_{\rho}^2\over \hat{m}_{\rho}\Gamma_{\rho}}
\right )\right ]^{-1},
\label{zdefn}\\
\tilde{T} &\equiv& 
\tilde{\Pi}_{\rho\omega}(m_{\rho}^2)/\hat{m}_{\rho}\Gamma_{\rho}
\label{TTwiddle}
\end{eqnarray}
with $\tilde{\Pi}_{\rho\omega}$ the real part of $\Pi_{\rho\omega}$ 
and $\hat{m}_V$ the real part of the complex pole position of the 
vector meson $V$.
The imaginary part of $\Pi_{\rho\omega}$ is given by 
$-G\hat{m}_\rho\Gamma_\rho$ with 
$G \equiv g_{\omega^I\pi\pi}/ g_{\rho^I\pi\pi}$.
In the limit that $\Gamma_\omega$ and $m_\rho^2-m_\omega^2$
are negligible, $z=1$, and the first term in Eq.~\ref{coupling} 
vanishes identically, making a determination of $g_{\omega^I\pi\pi}$ 
impossible in this limit. Fortunately, $z$ 
turns out to deviate sufficiently from $1$ to allow a 
reasonable determination of $g_{\omega^I\pi\pi}$.

With the above definitions, Eq.~\ref{ff1} becomes
\begin{equation}
F_{\pi}(s) = {f_{\rho\gamma}\over e}g_{\rho^I\pi\pi}\left[\tilde{P}_{\rho}(s) + 
|r_{\rm ex}| e^{i\phi_{e^+e^-}}((1-z)G-iz\tilde{T})\tilde{P}_{\omega}(s)\right] + {\rm background}
\label{ff2}\end{equation}
where ${f_{\omega\gamma}\over f_{\rho\gamma}} \equiv r_{ex}
\equiv |r_{\rm ex}| e^{i\phi_{e^+e^-}}$ and,
experimentally~\cite{blcp94},
\begin{equation}
|r_{\rm ex}| = \left[{\hat{m}_{\omega}^3\Gamma(\omega\to e^+e^-)\over
\hat{m}_{\rho}^3\Gamma(\rho\to e^+e^-)}\right ]^{1/2} = 0.296 \pm 0.005  
\end{equation}
using PDG07 \cite{pdg07} values for the widths and masses.
The full propagators appearing in Eq.~\ref{ff1} have been replaced in 
Eq.~\ref{ff2} with the same forms as were used in Sec.~\ref{models} 
(Eq.~\ref{propagator} for the KS and HLS models,
Eq.~\ref{GSprop} for the GS model, and Eq.~\ref{gpmodel} for the GP/CEN
family), but with their dimensionful numerators removed
(e.g., $\tilde{P}_\rho(s) = P_\rho(s)/m_\rho^2$ for the KS and HLS models).  
(This preserves the usual dimensionful definition of $f_{V\gamma}$.)

Fits to experimental data provide only the magnitude of 
the $\omega$ contribution to $F_{\pi}(s)$ and its phase relative to 
the dominant $\rho$ term, i.e., the modulus and phase of the coefficient of 
$\tilde{P}_\omega(s)$ inside the
square brackets in Eq.~\ref{ff2}. This coefficient is written compactly as 
$Ae^{i\phi}$ with $\phi$ the so-called Orsay phase.
It follows from the above that the values of $A$ and $\phi$ obtained from 
a given data set depend on the precise form of the $\rho$ propagator in the 
model used to represent $F_{\pi}(s)$.
Any separation of $G$ and $\tilde{T}$ is thus unavoidably model-dependent. 

\subsection{Numerical Results}

The Orsay amplitude and phase corresponding to the data sets and models 
described in sections \ref{models} and \ref{data} 
are shown in the first two rows of Tables 
\ref{table5} to \ref{table8}.  The third and fourth rows of 
these tables also show the real and imaginary parts of $z$, which are
determined in each case by the corresponding fit values of the $\rho$
mass and width. 
The range of Orsay phases found here is consistent within errors
with the old value of
$\phi = 116.7^\circ\pm 5.8^\circ$~\cite{blcp94} obtained 
using less precise data and a model of $F_{\pi}(s)$ similar to the KS 
model but with a constant background term instead of the 
$\rho^{\prime}$ resonance. 
The central value of our result for the phase, obtained as described 
previously, is then 
\begin{equation}
\phi = 113^\circ \pm 4^\circ_{model} \pm 2^\circ_{data}
\label{Orsayphase}
\end{equation} 
slightly lower than the result of Ref.~\cite{blcp94}, but
consistent with it within errors.  Omission of the KLOE data makes no
significant difference to this determination.
 
The Orsay phase uncertainty for any single model ranges from
2$^\circ$ for the high-precision SND data to 5$^\circ$ for the 
KLOE(02) data.  
As expected, the phase determination is slightly less precise
for the CMD-2(98) data than for the lower statistics CMD-2(94) 
data because
of the former's larger correlated systematic errors.
The precision of the phase determination from the new
data, for a given model, has been improved by a factor of $\sim 1.5-2$ 
compared to that from the earlier data.

\begin{figure}[ht]
\epsfig{figure=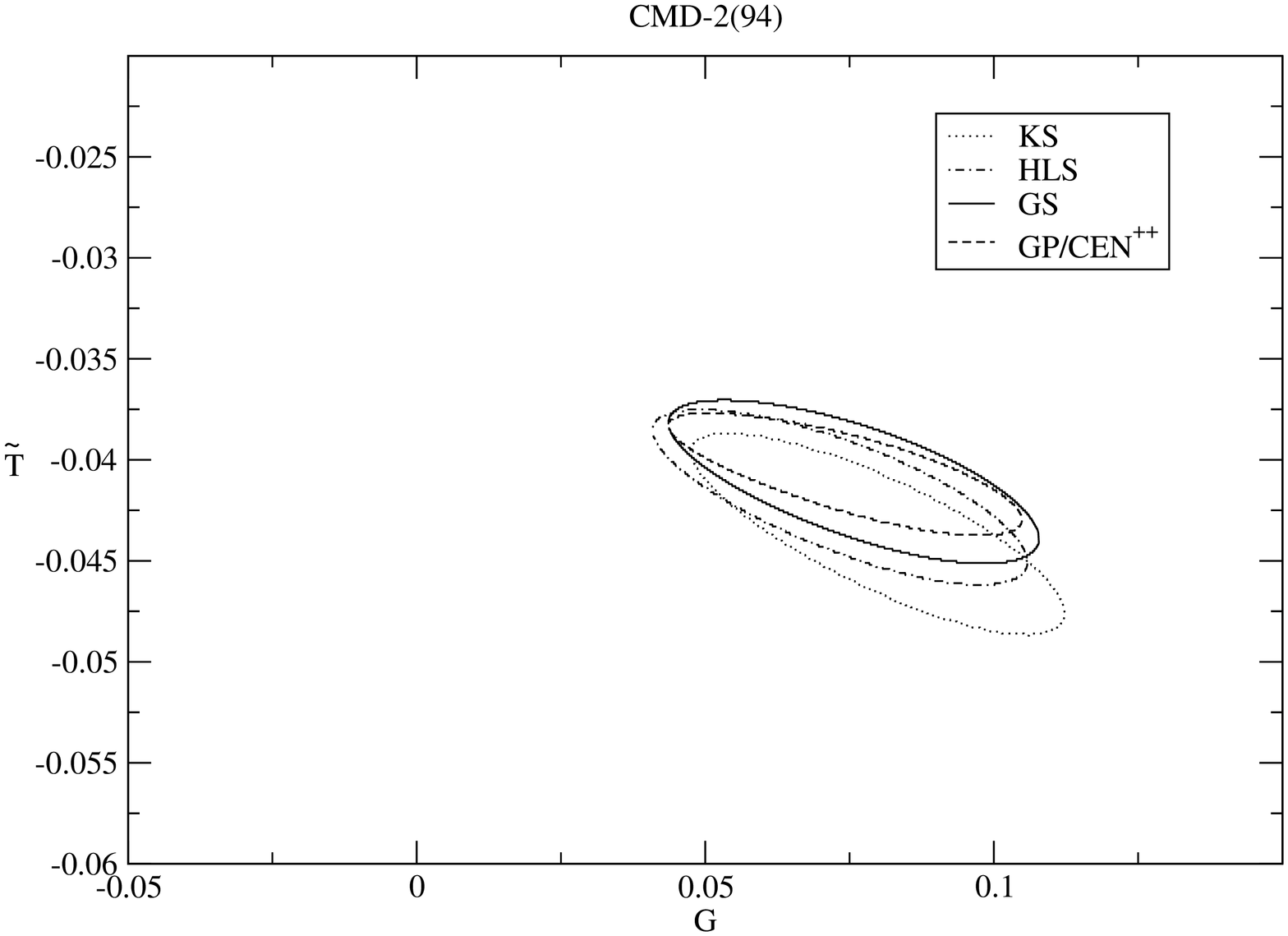, scale=0.50, angle=-90}
\caption{\label{CMD294MOW}CMD-2(94) $G$ and $\tilde{T}$ one-sigma regions.}
\end{figure}

\begin{figure}[ht]
\epsfig{figure=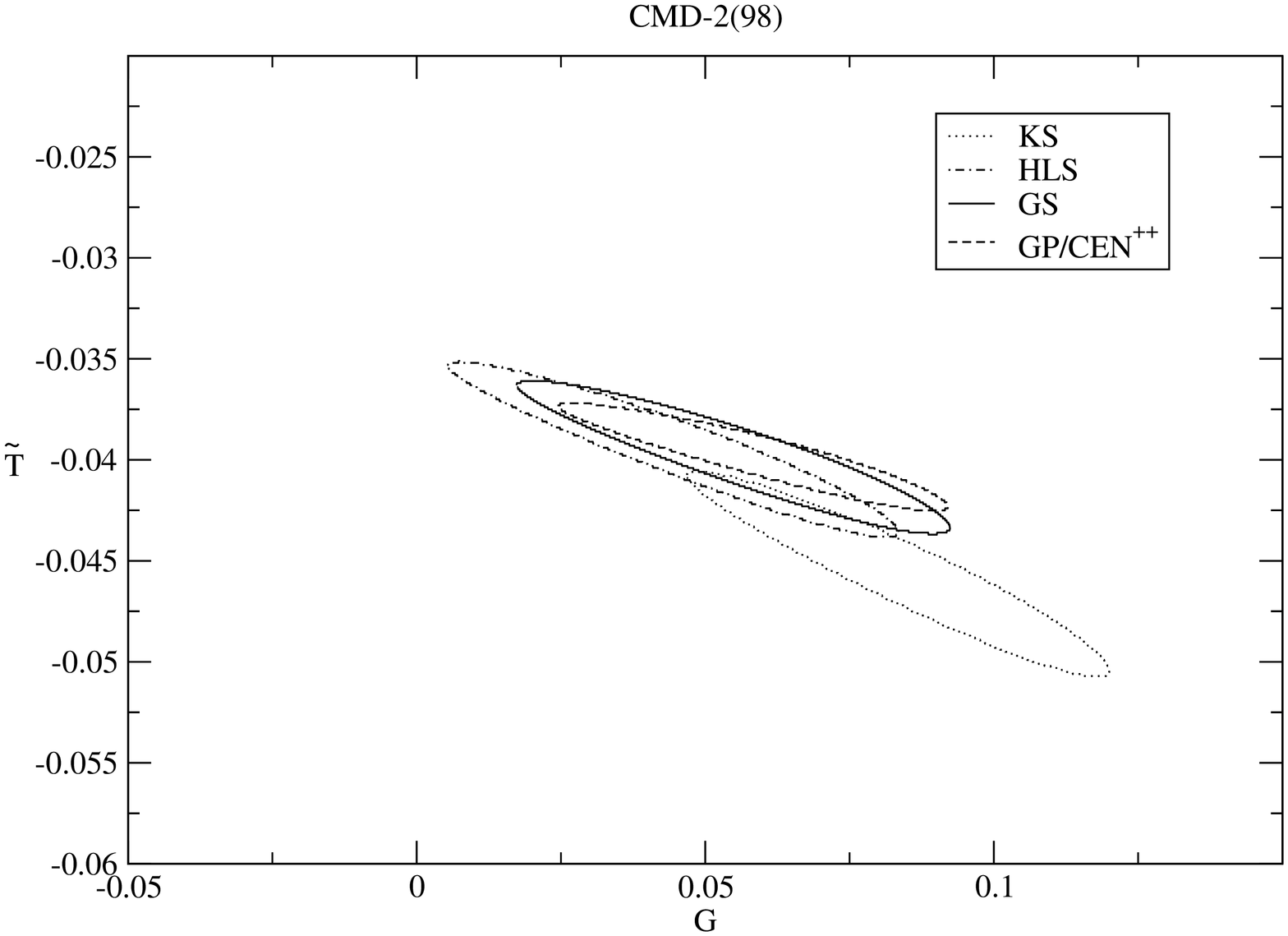, scale=0.50, angle=-90}
\caption{\label{CMD298MOW}CMD-2(98) $G$ and $\tilde{T}$ one-sigma regions.}
\end{figure}

\begin{figure}[ht]
\epsfig{figure=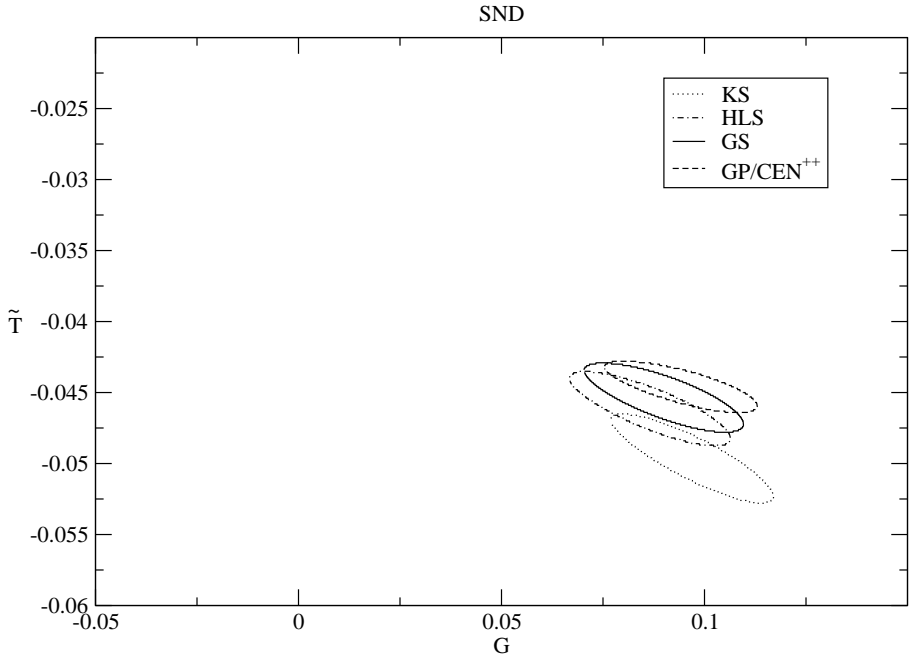, scale=0.50, angle=-90}
\caption{\label{SNDMOW}SND $G$ and $\tilde{T}$ one-sigma regions.}
\end{figure}

\begin{figure}[ht]
\epsfig{figure=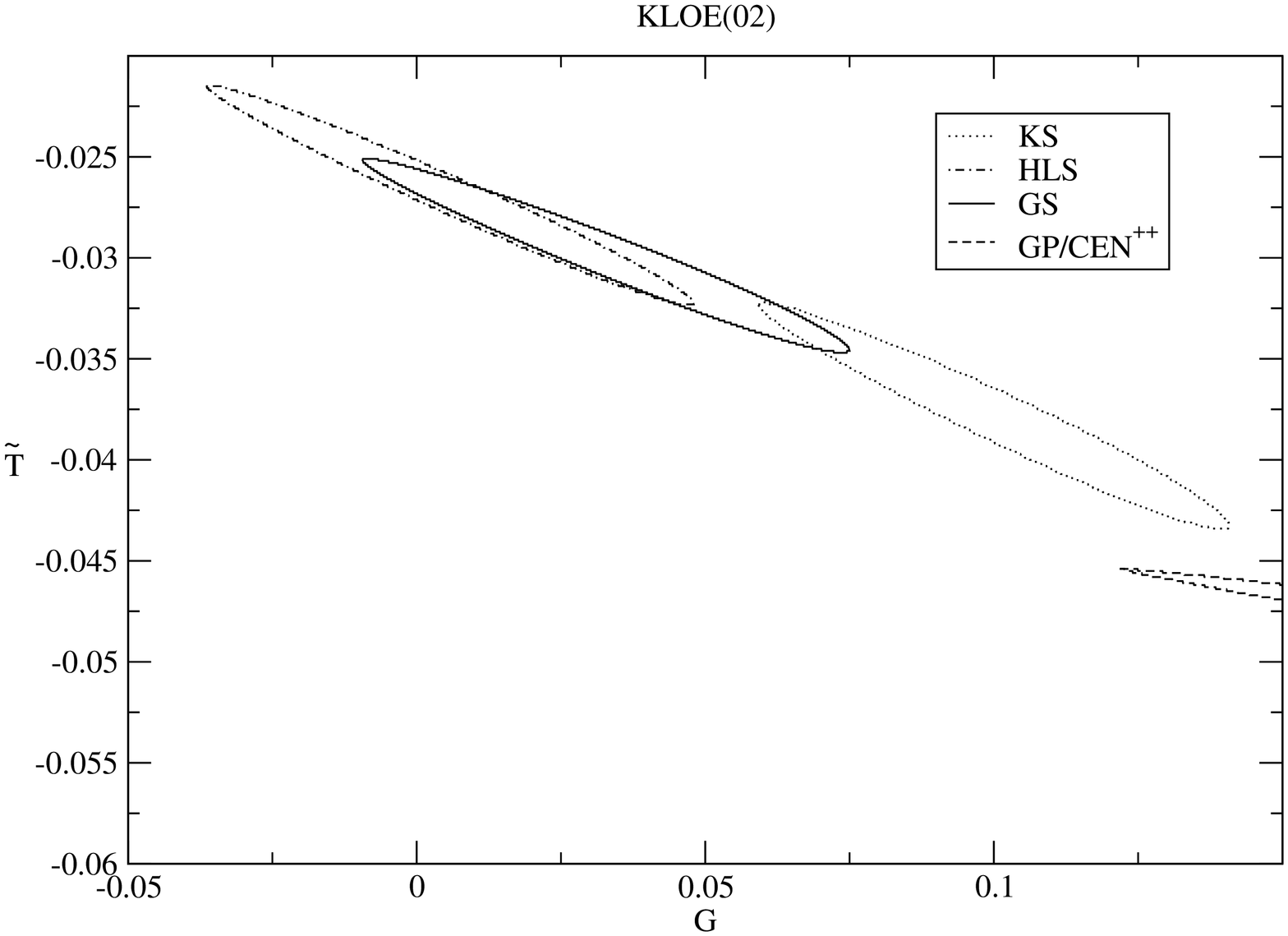, scale=0.50, angle=-90}
\caption{\label{KLOEMOW}KLOE $G$ and $\tilde{T}$ one-sigma regions.}
\end{figure}

The 30\% to 50\% reduction in phase uncertainty significantly improves
the reliability of the separation of pure mixing and direct 
$\omega\pi\pi$ coupling contributions to $F_{\pi}(s)$.
The results for $G$ and $\tilde{T}$ corresponding to the best fit 
values for $A$ and $\phi$ are shown in rows 5 and 6 of Tables~\ref{table5} 
to \ref{table8}. The one-sigma region in the $G$--$\tilde{T}$ space, shown for 
each data set and model in Figures~\ref{CMD294MOW}, \ref{CMD298MOW}, 
\ref{SNDMOW}, and \ref{KLOEMOW}, defines the uncertainties in $G$ and 
$\tilde{T}$ and is based on full propagation of the fit parameter 
covariance matrix, itself a reflection of the experimental covariances.  

\begin{table}
\caption{\label{table5}
Orsay amplitude, phase, and separated mixing and direct $\omega\pi\pi$
coupling parameters for the CMD-2(94) data.}
\vskip .1in
\begin{tabular}{|c||c|c|c|c|c|}
\hline
Parameter & KS & HLS & GS & GP/CEN$^+$ & GP/CEN$^{++}$ \\
\hline
$A$  & $0.0101 \pm 0.0008$ & $0.0103 \pm 0.0008$ & $0.0103 \pm 0.0008$ & $0.0107 \pm 0.0008$ & $0.0107 \pm 0.0005$\\
$\phi$ (deg) & $114 \pm 3$ & $112 \pm 3$ & $111 \pm 3$ & $110 \pm 3$ & $ 110 \pm 3$ \\
${\rm Re}(z)$  & $1.041\pm 0.003 $ & $1.046 \pm 0.002$ & $1.048\pm 0.002$ &$1.052\pm 0.002$ & $1.053 \pm 0.002$ \\
${\rm Im}(z)$  & $0.158 \pm 0.009$ & $0.133\pm 0.009$ & $0.121\pm 0.009$ &$0.102\pm 0.009$ & $0.103 \pm 0.009$ \\
\hline
$G$ & $0.080\pm 0.032$ & $0.073\pm 0.032$ & $0.076\pm 0.032$ & $0.074\pm 0.032$ & $0.074\pm 0.031$ \\
$\tilde{T}$ & $-0.043\pm 0.005$ & $-0.042\pm 0.004$ & $-0.041\pm 0.004$ & $-0.041\pm 0.003$ & $-0.041\pm 0.003$ \\
$\sin\phi_{\rm e^+e^-}$ & $0.120\pm 0.004$ & $0.124\pm 0.003$ & $0.124\pm 0.003$ & $0.129\pm 0.003$ & $0.129\pm 0.003$ \\
\hline
\end{tabular}
\end{table}

\begin{table}
\caption{\label{table6}
Orsay amplitude, phase, and separated mixing and direct $\omega\pi\pi$
coupling parameters for the CMD-2(98) data.}
\vskip .1in
\begin{tabular}{|c||c|c|c|c|c|}
\hline
Parameter & KS & HLS & GS & GP/CEN$^+$ & GP/CEN$^{++}$ \\
\hline
$A$  & $0.0107 \pm 0.0004$ & $0.0107 \pm 0.0004$ & $0.0107 \pm 0.0004$ & $0.0111 \pm 0.0004$ & $0.0111 \pm 0.0002$ \\
$\phi$ (deg) & $114 \pm 4$ & $109 \pm 3$ & $109 \pm 4$ & $108 \pm 4$ & $108 \pm 3$ \\
${\rm Re}(z)$  & $1.041\pm 0.002 $ & $1.046 \pm 0.002$ & $1.049\pm 0.002$ &$1.054\pm 0.001$ & $1.055 \pm 0.001$ \\
${\rm Im}(z)$  & $0.153 \pm 0.008$ & $0.128\pm 0.008$ & $0.116\pm 0.008$ &$0.087\pm 0.008$ & $0.087 \pm 0.008$ \\
\hline
$G$ & $0.083\pm 0.037$ & $0.044\pm 0.040$ & $0.055\pm 0.037$ & $0.058\pm 0.036$ & $0.058\pm 0.034$ \\
$\tilde{T}$ & $-0.046\pm 0.005$ & $-0.039\pm 0.004$ & $-0.040\pm 0.004$ & $-0.040\pm 0.003$ & $-0.040\pm 0.003$ \\
$\sin\phi_{\rm e^+e^-}$ & $0.127\pm 0.003$ & $0.130\pm 0.002$ & $0.130\pm 0.003$ & $0.135\pm 0.003$ & $0.135\pm 0.003$ \\
\hline
\end{tabular}
\end{table}

\begin{table}
\caption{\label{table7}
Orsay amplitude, phase, and separated mixing and direct $\omega\pi\pi$
coupling parameters for for the SND06 data.}
\vskip .1in
\begin{tabular}{|c||c|c|c|c|c|}
\hline
Parameter & KS & HLS & GS & GP/CEN$^+$ & GP/CEN$^{++}$ \\
\hline
$A$  & $0.0110 \pm 0.0004$ & $0.0112 \pm 0.0004$ & $0.0111 \pm 0.0004$ & $0.0116 \pm 0.0004$ & $0.0116 \pm 0.0002$ \\
$\phi$ (deg) & $116 \pm 2$ & $113 \pm 2$ & $113 \pm 2$ & $112 \pm 2$ & $111 \pm 2$ \\
${\rm Re}(z)$  & $1.037\pm 0.002 $ & $1.044 \pm 0.002$ & $1.047\pm 0.002$ &$1.053\pm 0.001$ & $1.053 \pm 0.001$ \\
${\rm Im}(z)$  & $0.166 \pm 0.007$ & $0.136\pm 0.007$ & $0.124\pm 0.007$ &$0.094\pm 0.007$ & $0.094\pm 0.007$ \\
\hline
$G$ & $0.097\pm 0.020$ & $0.087\pm 0.020$ & $0.090\pm 0.020$ & $0.093\pm 0.020$ & $0.094\pm 0.019$ \\
$\tilde{T}$ & $-0.050\pm 0.003$ & $-0.046\pm 0.003$ & $-0.045\pm 0.002$ & $-0.044\pm 0.002$ & $-0.044\pm 0.002$ \\
$\sin\phi_{\rm e^+e^-}$ & $0.130\pm 0.003$ & $0.133\pm 0.003$ & $0.133\pm 0.003$ & $0.140\pm 0.003$ & $0.140\pm 0.003$ \\
\hline
\end{tabular}
\end{table}

\begin{table}
\caption{\label{table8}
Orsay amplitude, phase, and separated mixing and direct $\omega\pi\pi$
coupling parameters for the KLOE(02) data. }
\vskip .1in
\begin{tabular}{|c||c|c|c|c|c|}
\hline
Parameter & KS & HLS & GS & GP/CEN$^+$ & GP/CEN$^{++}$ \\
\hline
$A$  & $0.0071 \pm 0.0003$ & $0.0081 \pm 0.0003$ & $0.0081 \pm 0.0003$ & $0.0111 \pm 0.0003$ & $0.0111 \pm 0.0003$ \\
$\phi$ (deg) & $119 \pm 6$ & $104 \pm 5$ & $107 \pm 5$ & $118 \pm 5$ & $119 \pm 5$ \\
${\rm Re}(z)$  & $1.034\pm 0.001 $ & $1.042 \pm 0.001$ & $1.046\pm 0.001$ &$1.055\pm 0.0006$ & $1.055 \pm 0.0006$ \\
${\rm Im}(z)$  & $0.173 \pm 0.004$ & $0.141\pm 0.004$ & $0.129\pm 0.004$ &$0.087\pm 0.003$ & $0.088 \pm 0.009$ \\
\hline
$G$ & $0.100\pm 0.041$ & $0.006\pm 0.042$ & $0.028\pm 0.042$ & $0.169\pm 0.054$ & $0.173\pm 0.050$ \\
$\tilde{T}$ & $-0.038\pm 0.006$ & $-0.027\pm 0.005$ & $-0.030\pm 0.005$ & $-0.047\pm 0.004$ & $-0.047\pm 0.002$ \\
$\sin\phi_{\rm e^+e^-}$ & $0.080\pm 0.002$ & $0.100\pm 0.001$ & $0.099\pm 0.001$ & $0.129\pm 0.003$ & $0.128\pm 0.003$ \\
\hline
\end{tabular}
\end{table}

As Figures \ref{CMD294MOW}--\ref{KLOEMOW} show, 
it is possible to constrain $\tilde{T}$ rather well. 
With the prescription above for handling averages and uncertainties,
we obtain
$$\tilde{T} = -0.044 \pm 0.006_{model} \pm 0.002_{data}.$$
The consistency across different models for each data
set is actually quite good.

The determination of $G$ is less precise, with the prescription
above for determining the average across models and
model-dependence uncertainty yielding
$$G = 0.080\pm 0.026_{model} \pm 0.015_{data}.$$  
For comparison, the uncorrected SND analysis \cite{snd05pipi}, using a
model very similar in form to the KS model described in 
Section~\ref{models}, reported
$G = 0.11\pm0.01$, consistent with our result, $G=0.10\pm 0.02$,
for the SND+KS case.
The present analysis favors a non-vanishing $G$, 
as expected on general grounds, with a vanishing $G$ lying $\sim 2\sigma$
away from our central value. 
An improved determination of
the Orsay phase, particularly in the case of the KLOE data,
is required to make further progress on this issue. To illustrate this point,
and also because of the lingering KLOE shape discrepancy \cite{kloefootnote}, 
if one omits the KLOE data
from the experimental averages one obtains results with a slightly 
reduced model-dependence uncertainty, 
$$\tilde{T} = -0.045\pm 0.004_{model}\pm 0.002_{data}$$
and
$$G = 0.084\pm 0.022_{model}\pm 0.016_{data}.$$

The values for $G$ and $\tilde{T}$ obtained in Tables 
\ref{table5}--\ref{table8} imply, using the earlier definitions,
and our prescription for combining the results corresponding
to different models, the result 
\begin{equation}
\tilde{\Pi}_{\rho\omega}(m_{\rho}^2) =
-4950\pm 530_{model}\pm 250_{data}\ {\rm MeV}^2
\label{repiroresult}\end{equation}
for the real part of $\Pi_{\rho\omega}(m_{\rho}^2)$.
The imaginary part of $\Pi_{\rho\omega}(m_{\rho}^2)$ is,
similarly, found to be $-8600\pm 2510_{model}\pm 1690_{data}\ {\rm MeV}^2$.
The result
is an estimate of the off-diagonal part of the physical $\rho-\omega$
self-energy matrix {\it at the $\rho$ mass} 
\begin{equation}
\Pi_{\rho\omega}(m_{\rho}^2) = 
(-4950\pm 780) + (-8600\pm 4200)i\;{\rm MeV}^2,
\label{selfenergy}
\end{equation}
where we have combined the model-dependence 
and experimental errors {\it linearly}.
Modestly reduced uncertainties are again obtained if the KLOE(02) data is 
excluded in forming the combined result. Explicitly:
\begin{equation}
\Pi_{\rho\omega}(m_{\rho}^2) =
(-4950\pm 670) + (-8890\pm 4000)i\;{\rm MeV}^2. 
\label{selfenergynokloe}
\end{equation}
This quantity is not to be confused with the 
sometimes quoted `effective' 
$\rho-\omega$ mixing matrix element, which contains
direct $\omega\rightarrow\pi\pi$ contributions.

The separation described here is possible only because the two 
contributions enter the form factor with different (but known) phases, 
as shown by Eq.~\ref{ff2}.  
Experimental determination of the combined phase (the Orsay phase)
then allows the magnitude of each contribution to be determined separately.
Any improvement in the separation of the
$\rho-\omega$ mixing and direct $\omega\rightarrow\pi\pi$
IB contributions thus requires corresponding improvements in the
determination of the Orsay phase.

\section{\label{conclusions} Conclusions}

Recent experiments 
provide a much-improved characterization of the
prominent $\rho-\omega$ interference shoulder in 
the $e^+e^-\rightarrow\pi^+\pi^-$ cross-section, making 
precision analyses of isospin breaking in the $\rho-\omega$ system possible.
Such an analysis is inherently dependent on the model 
chosen to represent the broad $\rho$ resonance contribution.  
By studying a range of models, all of which provide good fits to 
the data, we have shown that the model sensitivity of most fit 
parameters is generally
greater than the statistical uncertainty associated with any given model.  
The only exception is the phase, ${\rm Arg}(\delta)$, of the
parameter describing the strength of the $\omega$ contribution, 
for which the statistical uncertainty tends to be 
greater than the model sensitivity.

Any model sensitivity of course propagates to all quantities 
obtained from the fit parameters. The example discussed in 
Section~\ref{gminus2ib} is that of the $\rho-\omega$ 
interference contribution to the IB correction,
$\left[\delta a_{\mu}\right]_{had}^{LO}$,
required for incorporating hadronic $\tau$ decay data
into the determination of the SM expectation for $[a_{\mu}]_{had}^{LO}$.  
Using our result, Eq.~\ref{CVall}, the 
full IB correction obtained in Ref.~\cite{dehz02} using 
the method of Ref.~\cite{cen02} (see also~\cite{footnote10}),
$\left[\delta a_{\mu}\right]^{LO}_{had} = (-13.8 \pm 2.4)\times 10^{-10}$ 
(or $\left[\delta a_{\mu}\right]^{LO}_{had} = (-16.6 \pm 1.6)\times 10^{-10}$
based on the recent analysis of Ref.~\cite{davieretal09}),
receives only a small shift of $-0.4\times10^{-10}$
($+0.7\times 10^{-10}$) in its central value.
The uncertainty in the size of the correction, however, increases by a 
much more significant $1.5\times 10^{-10}$.
In arriving at the combined error, we have added the component associated with
the model-dependence of the $\rho-\omega$ mixing 
contribution {\it linearly} to the other errors,
for the reason discussed above.
While the increased error does not serve to resolve the
$\sim(17 \pm 10)\times 10^{-10}$~\cite{passera07, jegerlehner09}
($\sim(11\pm 5)\times 10^{-10}$~\cite{davieretal09})
$\tau$ vs. EM discrepancy
which existed prior to the preliminary BaBar ISR results,
it is nonetheless important to take into account 
in assessing the uncertainty in the IB corrections which
must be applied if one wishes to use $\tau$ decay data in 
evaluating the SM contribution to $[a_{\mu}]^{LO}_{had}$.

The precision of the recent electroproduction data also allows new 
improved results for the (model-dependent) separation 
of contributions to $F_{\pi}(s)$ from 
$\rho-\omega$ mixing and direct (non-mixing) IB $\omega\to\pi\pi$
transitions. 
We estimate the ratio of the two-pion couplings of the isospin-pure 
$\omega$ and $\rho$ to be $G = 0.080\pm 0.041$
where the uncertainty is conservatively chosen to encompass the experimental 
error intervals of all of the models considered. 
Further improvements in the determination of $G$ may be possible in future.
As can be seen in Figures~\ref{CMD294MOW}--\ref{KLOEMOW}, the 
off-diagonal part of the vector meson self-energy matrix
evaluated at the $\rho$ mass is very well constrained by 
the analysis. The central value we obtain for the real part is given 
in Eq.~\ref{repiroresult}, where the uncertainty again conservatively 
encompasses the full ranges corresponding to all models
and all data sets. Our central value of $\tilde{\Pi}_{\rho\omega}(m_\rho^2)$
which is free of contamination from direct $\omega\rightarrow\pi\pi$ coupling 
contributions, and thus differs from the usually quoted `effective' 
$\rho-\omega$ mixing matrix element obtained by ignoring such a 
direct coupling, should be used in place of the 
`effective' matrix element results in meson-exchange models of
IB NN scattering.

Further improvement of the separation of direct and mixing 
contributions requires a reduction in the uncertainty of the
Orsay phase, which in turn requires a further increase in the precision of the 
$e^+e^-\to\pi^+\pi^-$ data in the $\rho-\omega$ interference 
region. This requires not only increased statistics but also very 
precise beam energy calibration such as that expected from VEPP-2000.  
\vskip .2in
{\it NOTE ADDED: Subsequent to the posting of this paper, the BaBar
collaboration posted its results for the $e^+e^-\rightarrow\pi^+\pi^-$
cross-sections, obtained using the ISR method~\cite{babarpipi09}.
We will report the results of the analysis discussed here, 
applied to that data set, as soon as the data becomes publicly available.}
\begin{acknowledgments}
We thank G. Venanzoni for clarification on some of the KLOE systematic
errors and Z. Zhang, C. Yuan, P. Wang and X. Mo for extensive
interchanges on the relation between our results and those of
Ref.~\cite{davieretal09}. 
CW thanks S. Eidelman and I. Logashenko for valuable discussions about 
the CMD-2(98) data.
KM would like to acknowledge the hospitality of the CSSM, University of
Adelaide, and the ongoing support of the Natural Sciences and Engineering 
Research Council of Canada.
\end{acknowledgments}



\begin{thebibliography}{99}
\bibitem{kloefootnote}
While estimates of $a_{\mu}$ based on the new data
are consistent with one another, a discrepancy still exists
between the KLOE \cite{kloe08} results for $F_{\pi}(s)$ 
and those obtained by CMD-2 \cite{cmd203, cmd207} and SND 
\cite{snd05pipi, snd06pipi} in the region above the $\rho$.
The fact that this discrepancy is not yet 
understood \cite{kloe08, passera04} strikes a cautionary note.
\bibitem{footnote2}Davier~\protect\cite{daviernote}, 
working with the GS model, 
has noted that the $m_\rho$ and $\Gamma_\rho$ values
obtained from fits to the bare and vacuum-polarization-dressed versions 
of the CMD-2 cross-sections differ by $\sim 1$ MeV. Similar shifts
also occur for the other models. These shifts produce corresponding
shifts in the $\rho-\omega$ interference contribution to 
$\left[\delta a_\mu\right]_{had}^{LO}$.
The versions associated with the fits to the bare cross-sections,
quoted in the text, lie higher than those
associated with the fit to the vacuum-polarization-dressed values 
by $(0.4\rightarrow 0.6)\times 10^{-10}$, depending on the model.
The relevant versions for comparison to the $\tau$ data are,
as explained by Davier, those obtained from the
fit to the bare cross-sections~\protect\cite{daviernote}.  These
are also the versions relevant to the determination of 
$[a_\mu]^{LO}_{had}$.
\bibitem{GSrhopfootnote} The older PDG average value for the width of the 
$\rho^{\prime}$ is used with the GS model to ease comparison with the CMD-2
analyses. The fits are
largely insensitive to the precise value of the $\rho^\prime$ width.
\bibitem{footnote10} When quoting the earlier central value and uncertainty for
$[\delta a_{\mu}]^{LO}_{had;mix}$, we use the result obtained in 
Ref.~\cite{dehz02}, based on the same formalism as elaborated in 
Ref.~\cite{cen02}, but with improved experimental inputs.
\end{thebibliography}
\end{document}